\documentclass{article}
\usepackage{spconf,amsmath,graphicx}

\title{Explanations for Automatic Speech Recognition}

\newcommand{\ajitha}[1]{\textbf{\color{red}AR:{#1}}}

\newcommand{\xiaoliang}[1]{\textbf{\color{blue}Xiaoliang:{#1}}}

\name{Xiaoliang Wu, Peter Bell, Ajitha Rajan
}
\address{School of Informatics, University of Edinburgh}
\usepackage{url}
\usepackage{algorithm}
\usepackage{algorithmicx}
\usepackage{algpseudocode}
\usepackage{verbatim}

\usepackage{colortbl}
\usepackage{multirow}

\usepackage{xcolor}

\begin{document}
%
\renewcommand{\algorithmicrequire}{\textbf{Input:}}  
\renewcommand{\algorithmicensure}{\textbf{Output:}}
\maketitle
\begin{abstract}
\vspace{-3pt}
We address quality assessment for neural network based ASR by providing explanations that help increase our understanding of the system and ultimately help build trust in the system. 
Compared to simple classification labels, explaining transcriptions is more challenging as judging their correctness is not straightforward and transcriptions as a variable-length sequence is not handled by existing interpretable machine learning models. 

We provide an explanation for an ASR transcription as a subset of audio frames that is both a minimal and sufficient cause of the transcription. To do this, we  adapt existing explainable AI (XAI) techniques from image classification - (1) \texttt{Statistical Fault Localisation(SFL)}~\cite{SFLexplanation} and (2) \texttt{Causal}~\cite{DC-causal}. Additionally, we use an adapted version of Local Interpretable Model-Agnostic Explanations (\texttt{LIME})~\cite{LIME} for ASR as a baseline in our experiments. We evaluate the quality of the explanations generated by the proposed techniques 
over three different ASR -- Google API~\cite{Google}, the baseline model of Sphinx~\cite{Sphinx}, Deepspeech~\cite{deepspeech2014} -- and 100 audio samples from the Commonvoice dataset~\cite{commonvoice:2020}. 

\end{abstract}


\section{Introduction}
\label{sec:intro}
\vspace{-5pt}
Owing to the prevalence of ASR in our daily lives, concerns such as their quality and accountability have become particularly important. 
The complex and often blackbox~\cite{kastanos2020confidence} (inaccessible internal state) nature of neural network- based ASR makes it hard to ensure their quality.  We focus on addressing quality assessment of ASRs by providing explanations for a given transcription that can help increase our understanding of the ASR and have the  potential to help fix the failure causing faults and facilitate the accountability process. This is supported by Madsen et al.~\cite{madsen2021post} who state, ``\emph{
Quality assessment should be done through model explanations.}" 

XAI techniques\footnote{Our focus is restricted to \emph{post-hoc} explainable methods that provide their explanation \emph{after} a model is trained.} have emerged rapidly in the past five years. 
One of the first XAI techniques was developed for explaining image classification labels. It  was a local perturbation-based technique called LIME~\cite{LIME} that 
randomly samples inputs around the
instance being explained using perturbations of the original input instance. By
feeding these samples into the model, LIME creates a new, local and linear interpretable
model around the original input that can help explain the input/output instance.  Other popular explanation techniques include some variants of LIME~\cite{DLIME,LIMEtree,k-LIME}; SHAP~\cite{SHAP}, a game-theory based input feature ranking based on its contribution to the output prediction; DeepLIFT~\cite{deeplift} and Integrated Gradients~\cite{intergratedgradients}, gradient-based methods using a reference
input  (chosen by the user) and difference in original and reference activations backpropagated to the input layer for contribution scores of input features. 
Minimal Contrastive Editing (MICE) [37] is a perturbation and example-based explanation
method especially developed for NLP models. MICE describes contrastive
explanations as answers to questions in the form ``Why $p$ and not $q$?", why a prediction event $p$
happened instead of contrast event $q$. 
The majority of these XAI techniques attempt to explain prediction labels for image or natural language classification models. 

There are a few recent techniques that interpret speech-input models from a white-box perspective.
Specifically,~\cite{2018NeuronAP} introduced time-independent Neuron Activation Profiles (NAPs) for each activation neuron of a given ASR for certain groups of inputs. They discovered different layers will be activated by different types of inputs by clustering NAPs. Later,~\cite{2021Zhou} found that averaging the feature maps after the ReLU activation in each transposed convolutional layer produces interpretable time series data that summarizes encodings of features of speech at the corresponding layer. This also allowed them to successfully understand what properties of speech will be encoded in each layer. 
These techniques have two limitations: (1) They are proposed for a specific white-box model and neural network architecture, making it difficult to generalize to other models; and (2) They do not explain the transcription of ASR, making it difficult to assess its quality with these techniques.  
Compared to simple classification labels, transcriptions -- as a variable-length output sequence -- are more challenging for two primary reasons: (1)  existing interpretable machine learning models, like linear regression model that LIME is based on, cannot handle variable length output sequences, s
; and (2) Unlike classification labels that involve simple comparisons for determining their correctness, judging transcription correctness involves semantic comparisons with room for small deviations. 


 To explain ASR transcriptions, we first come up with a means to label transcriptions as Correct or Incorrect based on the degree of similarity to the desired transcription. We then aim to provide an explanation for an ASR transcription from a given input audio as a subset of audio frames. We refer to \emph{frames} in explanations as raw data bins in time dimension, and they are not to be confused with frames represented by features like filterbanks or MFCCs in the frequency or cepstral domains.  To provide explanations, we  adapt existing XAI techniques from image classification - (1) Statistical Fault Localisation (SFL)~\cite{SFLexplanation},  (2) \texttt{Causal}~\cite{DC-causal}. We choose to adapt \texttt{SFL} and \texttt{Causal}  since they are causality-based XAI techniques that produce minimal and sufficient explanations outperforming other state-of-the art techniques in the image classification domain. 
 Additionally, we use \texttt{LIME}, a popular XAI technique in the literature, as a baseline in our comparison. 
It is worth noting that we choose to adapt  XAI techniques from image classification rather than NLP classification tasks as segmentation of input text often produces words with specific meanings. However, in image- and speech-based tasks, input segmentation results in pixels or frames that have no meaning independently and only make sense when they are considered in logical groups. This similarity motivates us to choose image-based explanation techniques over NLP.

We evaluate quality of explanations from different techniques 
in terms of their size and consistency across different ASR.
We found \texttt{SFL} and \texttt{Causal} explanations did well on all three ASR systems with respect to size and consistency. 

\noindent Source code for \texttt{X-ASR} and examples from our experiment are available at ~\url{https://anonymous.4open.science/r/Xasr-6E11}. 

\section{Methodology}
\label{sec: methodlogy}
\vspace{-7pt}
Our framework, \texttt{X-ASR}, generates explanations for ASR transcriptions in two steps, 
(1) \emph{Classify} ASR transcriptions with respect to reference and (2) \emph{Adapt} existing image explanation techniques -- \texttt{SFL}, \texttt{Causal} and \texttt{LIME} -- to work over audio input. We need the classification means in Step 1 because perturbation-based explanations, considered in Step 2, rely on classifying outputs from input perturbations and use the changes in outputs to assign importance to perturbed input regions. The classification step relies on existing similarity metrics in the literature. Adapting image-based explanations for ASR in step 2 has not been explored previously. 
%

\subsection{Classifying ASR Transcriptions}
\label{sec:classify-ASR}
Unlike classification tasks where it is straightforward to judge whether the output label is Correct (matches with expected label) or Incorrect, transcriptions from ASR are harder to assign a binary label as it may differ from expected transcription but still be Correct. For example, the expected transcription for an audio input may be ``I'd like an apple", while the actual ASR output is ``I like apple". We might still consider the ASR transcription as acceptable. Assessing the correctness of an ASR transcription is subject to human judgement and allowances for differences from expected transcription.  

To address this, we attach Correct or Incorrect labels to transcriptions based on the extent of similarity to expected transcriptions, using a user-defined threshold, $T$.  
If the similarity to original transcription is higher than the threshold $T$, then the perturbed audio transcription is marked as Correct, and Incorrect otherwise. 
We support \emph{two similarity metrics} in \texttt{X-ASR}: (1) \texttt{WER}, the widely used Word Error Rate metric that scores similarity between two sentences based on number of insertions, deletions and substitutions; and (2) \texttt{Bert}, where we compute  the cosine similarity between the semantic vectors for the original and perturbed audio transcriptions using Sentence Bert~\cite{bertsentence}.

\subsection{Adapting Image Explanations for ASR}

We discuss three image-based explanation techniques that we adapt for ASR. All three techniques rely on input perturbations. We classify outputs from perturbed inputs by measuring similarity to the original transcription.

\begin{algorithm}[pt]
\footnotesize
  \caption{Explanation using SFL for ASR} 
  \begin{algorithmic}[1]
    \Require
      $F$: ASR model;
      $x$: original audio;
      $M$: SFL thecnique measurement M;
      $metric$: similarity metric including \texttt{Bert} and \texttt{WER}
    \Ensure
       a subset of frames $P_{exp}$;
       \State $T(x)$ = mutants from $x$
       \For{each frame $f_i$ in $P$}
       \State calculate the $<a_{ep}^{i},a_{ef}^{i},a_{np}^{i},a_{np}^{i}>$ from $T(x)$ \label{li: SFLunvector}
       \State $value$=$M( <a_{ep}^{i},a_{ef}^{i},a_{np}^{i},a_{np}^{i}>)$\label{li: SFLunvalue}
       \State $ranking = $ frames in $P$ from high $value$ to low \label{li: SFLranking}
       \State $P_{exp}$ = $\phi$ \label{li: SFLexp_begin}
       \For{each frame $f_i$ in $ranking$}
       \State $P_{exp}$ = $P_{exp}\cup \left \{ f_{i} \right \}$
       \State $x_{exp}$ = mask frames of $x$ that are not in $P_{exp}$ 
       \State $Flag$=$Similarity(F(x),F(x_{exp}),metric)$\footnotemark 
       \If{$Flag = Correct $}
       \State \Return{$P_{exp}$}
       \EndIf
       \EndFor\label{li: SFLexp_end}
       \EndFor
  \end{algorithmic}
\label{Alg: SFL_explanation}
\end{algorithm}
\footnotetext{The input to the $Similarity$ function is the original transcription, the transcription of the mutant and the similarity metric.} 

\textbf{Adapting SFL Explanations}: \textbf{SFL} \cite{SFLexplanation}
adapted statistical fault localization  from software testing literature, to rank importance of pixels for a given classification label and built minimal explanations from this ranking. Statistical fault localization is traditionally used to rank program elements such as statements based on their likelihood of being failure causing faults. 

We describe step-wise how this idea is used in image-based \texttt{SFL} explanations to understand how we use it for speech. \\
\emph{Step 1:} Given an input image $x$ classified as $y$ by  a DNN, a series of mutants is generated by randomly setting a certain proportion of pixels to the background color (masked pixels) and classify each of these mutants. If the mutant classification label is $y$ (matching original). Then the mutant is a passing test, while a mutant whose classification is not $y$ is considered a failing test. To keep the number of passing and failing tests balanced, once a failing test is discovered,  the number of pixels to be masked for the next mutant is reduced  to make it more likely to be a passing test. Conversely, increase the number of pixels to be masked if a passing test is discovered. \\
\emph{Step 2:} Within each mutant, some pixels are consistent with the original input while others are masked. For each pixel in the input sample, count the number of mutants in which the pixels are masked versus unmasked and the output is changed versus unchanged. Based on the count, compute the importance score for every pixel using the chosen fault localisation measure, like Tarantula~\cite{tarantula}. \\
\emph{Step 3:} Starting from the top ranked pixel, add pixels in the rank list progressively to the explanation set, until classification of the explanation set matches the original image classification, $y$. This is returned as the explanation for an input image $x$ being classified as $y$. 

When applied to the audio signal, we use frames instead of pixels in all steps. We generate mutant audios by randomly selecting frames and setting the data points in them to zero.  We use similarity metrics \texttt{BERT} or \texttt{WER} for mutant classsification in Step 1 and explanation classification in Step 3. We show SFL explanations adapted for ASR in  Algorithm~\ref{Alg: SFL_explanation}. In Lines~\ref{li: SFLunvalue} and~\ref{li: SFLranking} in Algorithm~\ref{Alg: SFL_explanation}, for each frame $f_i$, we construct a parameter vector $ <a_{ep}^{i},a_{ef}^{i},a_{np}^{i},a_{np}^{i}>$ where $a_{ep}^{i}$ is the number of mutants in the mutants set labeled Correct in which $f_i$ is not masked; $a_{ef}^{i}$ is the number of mutants in the mutant set labeled Incorrect in which $f_i$ is not masked; $a_{np}^{i}$ is the number of mutants in the mutant set labeled Correct in which $f_i$ is masked; $a_{np}^{i}$ is the number of mutants in the mutant set labeled Incorrect in which $f_i$ is masked. After that, we use the ranking measurement in~\cite{SFLexplanation} to compute the importance and rank frames within the given audio. Finally, as shown in Line~\ref{li: SFLexp_begin} to Line~\ref{li: SFLexp_end} in Algorithm~\ref{Alg: SFL_explanation}, we create the explanation for a given audio following Step 3 of the method. 

\begin{algorithm}[pt]
\footnotesize
  \caption{Responsibility} 
  
  \begin{algorithmic}[1]
    \Require
      $P_i$: a partition;
      $F$: PR model;
      $x$: an original audio;
      $metric$: similarity metric including \texttt{Bert} and \texttt{WER}
    \Ensure
       a responsibility map of $P_i$;
       \State $X_i$ = the set of mutants obtained from $x$ by masking subsets of $P_i$
       \For{each superframe $P_{i,j}$ in $P_i$}
       
       \State{\footnotesize $\tilde{X_i^j}$ = $\left \{ x_{m} : P_{i,j}\ is\ not\ masked\ in\ x_{m}\ and\ x_m\ is\ Correct\right \}$}\label{li: ResXij}
       \State $k$ = $min\left \{diff(x_{m},x)|x_{m}\in \tilde{X_i^j} \right \}$\label{li: Resfindk}
       
       \State $r_{i,j}\ =\ \frac{1}{k+1}$\label{li: Resun}
       \EndFor
       
       \State 
       \Return{$r_{i,0},\cdots ,r_{i,\left | P_i \right |-1}$}
  \end{algorithmic}
\label{Alg: responsibility}
\end{algorithm}

\textbf{Adapting Causal for ASR:}  In 2021, a causal theory-based approach, \texttt{Causal}, was proposed for computing explanations of image classifier~\cite{DC-causal}. Explanations from \texttt{Causal} were robust in the presence of image occlusions and outperformed SOTA. 
Causation is a relation between two events A (the cause) and B (the
effect) when A causes B. Counterfactual theories~\cite{davidhume} to define causation in terms of a
counterfactual relation state that ``An event A causally depends on B if, and only if:
1. If B had occurred, then A would have occurred.
2. If B had not occurred, then A would not have occurred."

\texttt{Causal} uses a framework of causality proposed by~\cite{causality} which extends counterfactual
reasoning by considering contingencies which are defined as changes in the
current setting. \texttt{Causal} quantifies
causality and expresses the degree of responsibility for any actual cause defined as the
minimal change required to create a counterfactual dependence. The responsibility is
defined as $1/(1+k)$ where $k$ is the size of the smallest contingency. The responsibility
measure for an actual cause can take any value between 0 and 1, where 0 means no
causal dependency and 1 means very strong causal dependency.
The algorithm for image-based explanations is briefly outlined below: 
First, partition the original image, without any overlap, into blocks referred to as superpixels.  Since there are several ways to partition an image into blocks, many such partitioning methods are considered and the resulting superpixels. A superpixel $S$ is defined as the \emph{cause} of an  image $x$ being labelled as $y$ when there is a subset $P$ of superpixels of $x$ that satisfies the following three conditions, (1) the superpixel $S$ does not belong to $P$, (2) masking  any subset of $P$ (setting pixels within to background color) will not affect the label $y$, and (3) masking both $P$ and superpixel $S$ at the same time will change the classification. $P$ is referred to as the witness of superpixel $S$ and the size of the smallest witness (contingency to create a counterfactual dependence) is used to assign a responsibility for each superpixel. When the size of the smallest witness is smaller, the actual cause $S$ requires less changes in other superpixels to help it achieve a counterfactual dependence. That is, there is a stronger causal relationship between it and the image being correctly classified. 
The superpixels are then iteratively refined into smaller partitions, repeating the computation of responsibility. This process is repeated with several partitioning methods to come up with a ranked list of pixel responsibilities which is then used to construct explanations. 

When applied to the audio signal, we partition the original audio, without any overlap, into superframes. We then compute responsibility of superframes, , as shown in Algorithm~\ref{Alg: responsibility}, and iteratively partition them further, repeating responsibility computation. As shown in Lines~\ref{li: ResXij},~\ref{li: Resfindk} and~\ref{li: Resun} in Algorithm~\ref{Alg: responsibility}, responsibility of a superframe, $P_{i,j}$\footnote{$P_{i,j}$ means this is the $j^{th}$ superframe in the $i^{th}$ partition.}, is $r(i,j)=1/(k+1)$, where 
$k$ is equal to the minimum difference between a mutant audio and the original audio over mutants $x_m$ that do \emph{not} mask $P_{i,j}$ and are labeled Correct.  Masking a subset of superframes for computing witness in condition (2) involves setting a random selection of frames within to zero. 

\textbf{Adapting LIME Explanations:}
\texttt{LIME}, proposed in ~\cite{LIME}, is an XAI technique that can be applied to any model without needing any information about its structure. \texttt{LIME} provides a local explanation by replacing a complex neural network (NN) locally with something simpler, for example a linear regression model. \texttt{LIME} creates many perturbations of the original image by masking out random segments, and then weights these perturbations by their ‘closeness’ to the original image to ensure that drastic perturbations have little impact. It then uses the simpler model(for example, the linear model) to learn the mapping between the perturbations and any change in output label. This process allows \texttt{LIME} to determine which segments are most important to the classification decision.

When applied to the audio signal, we replace image segments with audio frames. 
It is worth noting that \texttt{LIME} is designed to output the importance ranking of pixels, or frames in our case,  and this ranking is considered a \texttt{LIME} explanation. Inspired from minimal and sufficient explanations in \texttt{SFL} and \texttt{Causal}, we construct smaller \texttt{LIME} explanations using a greedy approach that starts from the top ranked frame, and then adds frames in the rank list iteratively to the explanation until classification of the explanation is correct with respect to the original audio transcription. We use these potentially smaller \texttt{LIME} explanations in our experiments in Section~\ref{sec: results}




\section{Experiments}
\vspace{-7pt}
We evaluate the explanations generated for ASR models using three quality metrics that are described below. We use three different ASR -- Google API~\cite{Google} (referenced as Google in the results), baseline model of Sphinx~\cite{Sphinx} and Deepspeech~\cite{deepspeech2014}) 0.7.3 version -- and 100 audio samples from Commonvoice dataset~\cite{commonvoice:2020}. Within \texttt{X-ASR},  we evaluate three explanation techniques mentioned in Section~\ref{sec: methodlogy}, namely, \texttt{SFL, Causal, LIME}, with two similarity metrics (\texttt{Bert} and \texttt{WER}) used to classify transcriptions from perturbed inputs. We use the default setting for every ASR.
We use Google Colab Pro with two NVIDIA Tesla T4 GPUs (16GB RAM, 2560 cores)  to run our experiments. 
We use the following parameters in our experiments:
an audio sampling rate of $16000Hz$, frame length of $512$.
For SFL, Causal and LIME, the mutation factor is $0.05$ which determines the proportion of frames to be randomly selected for each mutation. The size of the mutants set is $100$. For Causal, we run the experiments $3$ times to get a reliable ranking of frames. 
For similarity metrics, classification threshold for \texttt{Bert} is $0.5$ and for \texttt{WER} is $0$. We choose a zero threshold for \texttt{WER} to emulate strict classification.  
Additionally, we investigate the effect of choosing other classification threshold values for the similarity metrics and report our findings in Section~\ref{sec: results}. 

\textbf{Quality Metrics for Explanations:} 
We use two previously used quality metrics from image classification explanations~\cite{metrics, SFLexplanation}  to compare and contrast explanations over three ASR, namely, \emph{size} and \emph{consistency}. We use size of the explanation as number of frames in the explanation versus the input. When size is smaller, the quality of the explanation is better as it indicates the technique is more selective in identify key frames important to the output. The consistency metric assesses the degree to which similar explanations are generated from different ASR for a given audio. We assess the consistency of explanation methods using Google API as the reference and calculate the fraction of frames that stay the same in explanations generated with the other systems, namely Sphinx and Deepspeech. 

\section{Results and Analysis}
\label{sec: results}
\vspace{-7pt}

\begin{figure}
\centering
\includegraphics[scale=0.65]{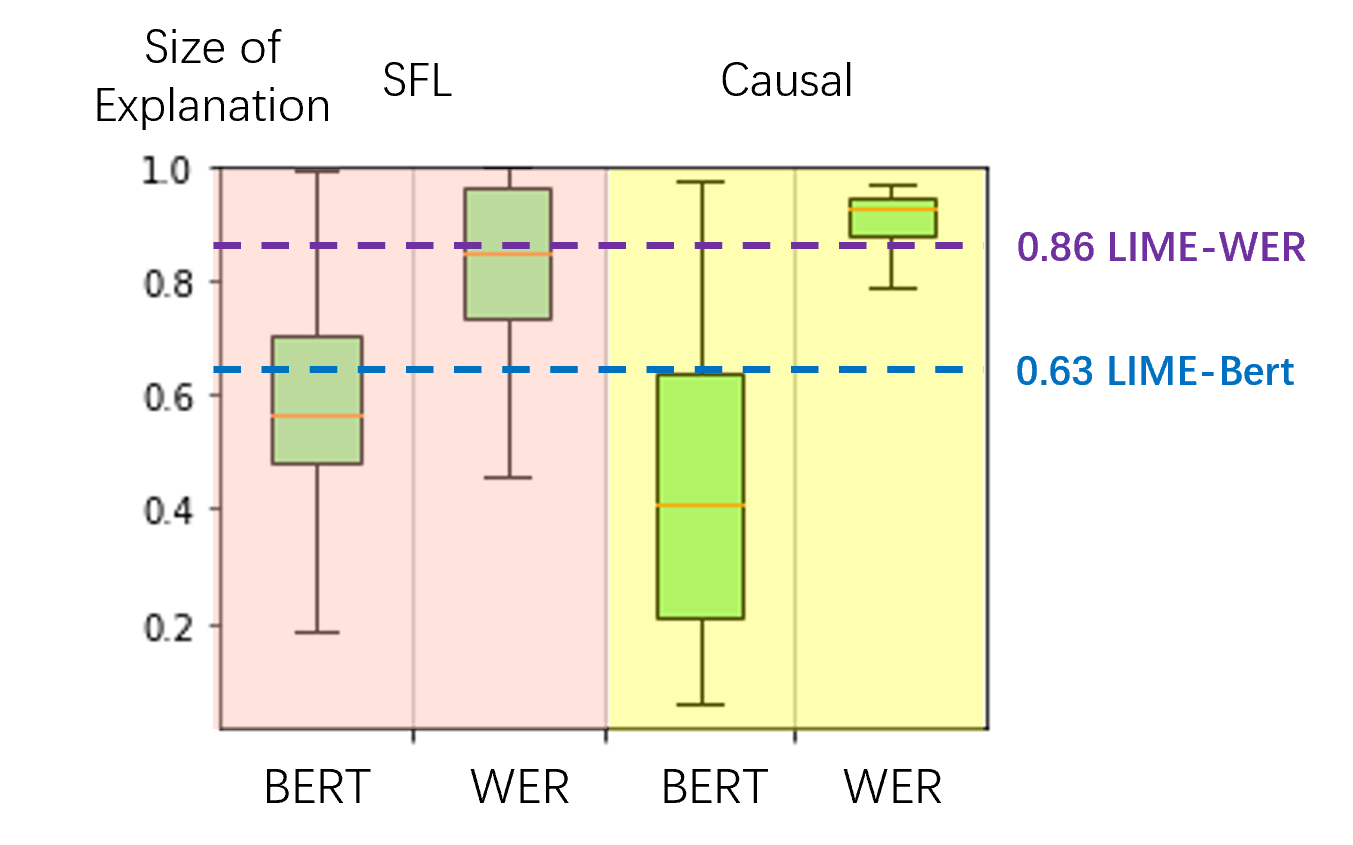}
\caption{Size of explanations using SFL and Causal against LIME using each of 2 similarities on Google ASR.}
\label{fig: Size}
\vspace{-5pt}
\end{figure}

\begin{table}[h]
\small
\centering
\begin{tabular}{|cc|c|c|c|}
\hline
\multicolumn{2}{|c|}{}                                                                                                                       & SFL                         & Causal                      & LIME                        \\ \hline
\rowcolor[HTML]{FFCCC9} 
\multicolumn{1}{|c|}{\cellcolor[HTML]{FFCCC9}}                                                                                        & Bert & {\color[HTML]{FE0000} 0.81} & 0.61                        & 0.80                        \\ \cline{2-5} 
\rowcolor[HTML]{FFCCC9} 
\multicolumn{1}{|c|}{\multirow{-2}{*}{\cellcolor[HTML]{FFCCC9}\begin{tabular}[c]{@{}c@{}}Consistency\\ (Google-Sphinx)\end{tabular}}} & WER  & 0.97                        & {\color[HTML]{FE0000} 0.98} & 0.92                        \\ \hline
\rowcolor[HTML]{FFFC9E} 
\multicolumn{1}{|c|}{\cellcolor[HTML]{FFFC9E}}                                                                                        & Bert & {\color[HTML]{FE0000} 0.67} & 0.60                        & {\color[HTML]{FE0000} 0.67} \\ \cline{2-5} 
\rowcolor[HTML]{FFFC9E} 
\multicolumn{1}{|c|}{\multirow{-2}{*}{\cellcolor[HTML]{FFFC9E}\begin{tabular}[c]{@{}c@{}}Consistency\\ (Google-Sphinx)\end{tabular}}} & WER  & 0.96                        & {\color[HTML]{FE0000} 0.98} & 0.89                        \\ \hline
\end{tabular}
\caption{Consistency(with respect to Sphinx or Deepspeech) of explanations generated by three explanation methods across two similarity metrics using Google ASR.
}
\label{tab: all}
\vspace{-5pt}
\end{table}

All three techniques in our method, \texttt{SFL, Causal, LIME}, were able to generate explanations for every input audio in our dataset and across all three ASR. 
Figure~\ref{fig: Size} shows size of explanations generated by the different techniques with the two similarity metrics on Google ASR. Results on other two ASR had a similar trend and are not shown here due to space limitations. 
We find SFL and Causal outperform the LIME counterpart for \texttt{Bert} by generating significantly smaller explanations (statistically significant difference confirmed with one-way Anova followed by post-hoc Tukey's test~\cite{tukey}). This is because both SFL and Causal not only pay attention to the changes in the output after masking a superframe but also take the number of other superframes affected by masking. Causal generates smaller explanations than SFL (statistically significant) owing to stricter conditions from causal-theory in assigning responsibility of a superframe and the iterative refinement of superframes.

Finally, there is a difference in size based on similarity metric. 
\texttt{WER} generates larger explanations than \texttt{Bert}. This is because \texttt{WER} has no tolerance for difference in transcription. As a result, more perturbations satisfy the condition for changing output classification. \texttt{Bert} accepts some changes in transcriptions with $0.5$ threshold. Raising the threshold for \texttt{WER} would also result in smaller explanations. Experiments with other thresholds in both metrics followed a similar trend. 
Table~\ref{tab: all} shows consistency of explanations with Google ASR versus Sphinx (rows highlighted in pink) and Google versus Deepspeech (rows highlighted in yellow). We find \texttt{SFL} is most consistent with \texttt{Bert} similarity on both ASR pairs. \texttt{Causal} is most consistent with \texttt{WER} similarity measure owing to its large explanation size ($92\%$ of input frames on average). 
This also follows for \texttt{LIME} explanations that are larger in size. 
Clearly, size and consistency are conflicting metrics and we believe \texttt{SFL} with \texttt{Bert} achieves a good compromise between the two -- $56\%$ size and $74\%$ consistency, on average.

\section{Conclusion}
\vspace{-7pt}
We propose the first framework, \texttt{X-ASR}, for generating explanations for ASR that supports three techniques: \texttt{SFL, Causal} and  \texttt{LIME}. 
We found  \texttt{SFL} and  \texttt{Causal} explanations perform comparably while outperforming LIME on size and consistency. 
The study presented only investigates perturbation-based explanation techniques that are portable and do not consider ASR structure. We plan to investigate white-box ASR explanation techniques and their effectiveness in the future. 


\bibliographystyle{IEEEbib}
\bibliography{refs}

\end{document}